\documentclass[a4page]{article}

\usepackage[pdftex]{color,graphicx,hyperref}
\usepackage{amsmath,amsthm,amsfonts,amssymb}

\usepackage{caption}
\usepackage{indentfirst}
\usepackage{authblk}
\usepackage{subfigure}
\usepackage{apacite}
\usepackage{natbib}

\newcommand*{\email}[1]{%
    \normalsize\href{mailto:#1}{#1}\par
    }

\newcommand{\Q}{{\mathbb{Q}}}

\newcommand{\E}{{\mathbb E}}

\newcommand{\1}{\mathbf{1}}

\theoremstyle{definition} 
\theoremstyle{definition} \newtheorem*{remark*}{Remark}

\usepackage{algpseudocode}
\usepackage[ruled,vlined]{algorithm2e}

\title{The importance of being scrambled: supercharged Quasi Monte Carlo}
\date{ 16/10/2023 }

\author{J. Hok}
\affil{Investec Bank, London, United Kingdom\\ \email{julienhok@yahoo.fr}}

\author{S. Kucherenko}
\affil{BRODA Ltd, London, United Kingdom\\ \email{s.kucherenko@broda.co.uk}}

\begin{document}
\maketitle

\textbf{Abstract:} {In many financial applications Quasi Monte Carlo (QMC) based on
Sobol’ low-discrepancy sequences (LDS) outperforms Monte Carlo
showing faster and more stable convergence.
However, unlike MC QMC lacks a practical error estimate. Randomized QMC (RQMC)
method combines the best of two methods.
Application of scrambled LDS allow to compute confidence intervals around the estimated value,
providing a practical error bound. Randomization of Sobol' LDS
by two methods: Owen's scrambling and digital shift are compared
considering computation of Asian options and Greeks
using hyperbolic local volatility model.
RQMC demonstrated the superior performance over standard QMC
showing increased convergence rates and providing
practical error bounds around the estimated values.
Efficiency of RQMC strongly depend on the scrambling methods. We recommend using
Sobol’ LDS with Owen’s scrambling. Application of effective dimension
reduction techniques such as the Brownian bridge or
PCA is critical to dramatically improve the efficiency of QMC and RQMC
methods based on Sobol’ LDS.}


\section{Introduction}


Monte Carlo (MC) is a unique universal method widely used in valuation
of complex financial instruments and risk management engines.
Its convergence rate $O( \frac{1}{\sqrt{N}})$ does not depend on the number of dimensions $d$
although it is rather slow. Here $N$ is the number of sampled points or the number of path.
It also provides practical error estimates through the computation of confidence intervals.

Unlike random numbers on which MC is based which are known to have bad uniformity properties,
deterministic low-discrepancy sequences (LDS) are designed to fill multidimensional
space as uniformly as possible. It results in the significantly improved  convergence rate
of the Quasi Monte Carlo (QMC) method based on LDS. Asymptotically,
it is $O( \frac{1}{N})$, which is much higher than
the convergence rate of MC. However, this theoretical estimate depends on the dimensionality of the problem and in practice the number of paths N required to achieve a given standard error may not be any lower than that of MC.
For problems in quantitative finance, dimensions $d$ can reach many thousands. It has led to a misconception
that QMC is not efficient in high dimensions. In practice, the effectiveness of QMC depends
not on the nominal dimension $d$ but on the so-called effective dimensions.
There are many problems for which the $d$-dimensional function $f$ is dominated by the first few variables
(low {\it{effective dimension in the truncation sense}}) or can be well
approximated by a sum of low-dimensional function in the ANOVA decomposition (low {\it{effective dimension in the superposition sense}}).
Such functions are very efficiently integrated by QMC achieving
the convergence rate close to $O( \frac{1}{N})$. It has been shown that
typically problems in finance have low effective dimensions or that effective dimensions can be reduced
by applying special sampling schemes.

One of the main drawbacks of the QMC methods is that since LDS are deterministic
there is no statistical method of computing the standard error of the estimate. It means that in particular there is no clear termination criterion for stopping simulation after reaching the required tolerance.
There are techniques, known under the name of randomized QMC (RQMC),
which introduce appropriate randomizations in the
construction of LDS. It allows for measuring integration errors through a
confidence interval similarly to MC while preserving and often
improving the convergence rate of QMC. While the superior performance of the QMC
methods based on Sobol LDS has been widely studied and used,
there is only a handful of pure academic papers in which RQMC methods were applied to problems
in finance. It was shown that RQMC offers both enhanced efficiency in comparison with pure QMC
and ability to produce confidence intervals, however it is still not widely used
in finance by practitioners. This work aims to bridge this gap.
We consider two popular methods of LDS randomisation: random digital shift and Owen's scrambling.

This paper is organized as follows: Section 2 provides a brief
review of MC, QMC and RQMC methods. In Section 3,
we introduce the ANOVA decomposition and the concept of effective dimensions.
The time-homogeneous hyperbolic local volatility model is presented in Section 4.
Two different time discretization schemes are considered in the next Section.
In Section 6, Monte Carlo simulation of option pricing and Greeks are discussed.
Section 7 presents numerical results. Finally,
the conclusions are given in the last Section.

\section{MC, QMC and RQMC methods}\label{sec:MC, QMC and RQMC methods}

The MC method solves a problem by simulating the underlying process and then
calculating the average result of the process. It can be formulated as computation
of the multidimensional  integral

\begin{equation}\label{multiDintegral}
I[f] = \int_{H^d} f(X)dX.
\end{equation}
Here function $f(X)$ is integrable in the $d$-dimensional unit hypercube $H^d$.
The MC quadrature formula is based on the probabilistic interpretation
of an integral as an expectation. The standard MC estimator of the expectation is

\begin{equation}\label{MCIntegral}
\hat\mu_N = \frac{1}{N} \sum_{i=1}^N f(X_i),
\end{equation}
where $\{X_i\}$ is a sequence of random points of length $N$ uniformly distributed in $H^d$. The approximation
$\hat\mu_N$ converges to $I[f]$ with probability $1$. An integration error according to the Central Limit Theorem is
$ \frac{\sigma(f)}{\sqrt{N}}$, where $\sigma^2(f)$ is the function variance.
Although typically $\sigma^2(f)$ is unknown, an unbiased estimate of it can be obtained
as well as confidence intervals (Table \ref{ComparionTable_MC_RQMC}).
The convergence rate of MC does not depend on the number of variables
$d$ but it is rather slow. It is known that random number sampling is prone to clustering.
As new points are added randomly, they do not necessarily fill the gaps between already sampled points.

In the classical Quasi-Monte Carlo (QMC) method independent random points  $\{X_i\}$
are replaced by a deterministic set of points such as LDS, which are
designed to cover the unit hypercube more uniformly than random points. Successive
LDS points “know” about the position of previously sampled points and “fill” the gaps between
them. The QMC algorithm for the evaluation of the integral (\ref{multiDintegral}) has a form similar to (\ref{MCIntegral})
where instead of random points $\{X_i\}$ LDS points $\{Q_i\}, Q_i \in H^d, i=1,...,N$ are used.
Sobol LDS also known as digital $(t, d)$ sequences in base 2 are
the most known and widely used LDS in finance due to their efficiency  \citep{glasserman2004monte} .

The efficiency of a particular Sobol’ LDS generator depends on the so-called direction numbers. In this
work we used BRODA's SobolSeq generator \citep{BRODA}.
Sobol' sequences produced by
BRODA's SobolSeq satisfy additional uniformity properties: Property $A$ for all dimensions
(currently maximum dimension $d=$131072) and Property $A'$ for adjacent dimensions.
It has been shown in \cite{SobAsoKreiKuch11} on a number of different tests 
that BRODA’s SobolSeq generators generally outperform other considered in the paper LDS generators.
These results were corroborated in other publications \citep{renzitti2020}.

A major drawback of the QMC method is the lack of practical estimates of the integration error.
A classical worst-case error bound for numerical integration by QMC is given by the Koksma-Hlawka inequality.
Although this bound can be used to get asymptotic convergence rates, it is too conservative and complex for computation
of practical error estimates.

Randomized QMC (RQMC) method combines the accuracy of QMC with the MC-type error estimation.
Consider a set of randomised replications $\{V_i\}$ of $\{Q_i\}$.
In the RQMC method (a) for a fixed $i$ each point $V_i$  is uniformly distributed $V_i \sim U[0,1]^d$;
(b) the point set $\{V_i\}, i=1,...,N$ is LDS with probability 1.

Consider a set of $K$ randomised replication $\{V_i\}={V_i^k}, k=1,...,K$. We denote
by $\hat\mu_n^k$ the $k-th$ RQMC estimator for (\ref{multiDintegral}):
\begin{equation}\label{RQMCIntegral}
\hat\mu_n^k = \frac{1}{n} \sum_{i=1}^n f(V_i^k),
\end{equation}
and by $\bar\mu_n$ the sample mean
\begin{equation}\label{RQMCAvIntegral}
\bar\mu_n = \frac{1}{K} \sum_{k=1}^K \hat\mu_n^k.
\end{equation}
We note that $\hat\mu_n^k$ are i.i.d., hence the sample standard deviation of this estimator and
the corresponding root mean square error (RMSE) can be computed in the same way as for MC. Table
\ref{ComparionTable_MC_RQMC} provides RMSE $\varepsilon_{MC}$ and $\varepsilon_{RQMC}$ for the MC and RQMC respectively.
It also provides the expressions for confidence intervals. It is assumed that $K$ is large enough so that the sample mean
$\bar\mu_n$ is normally distributed. $z_{\delta}$ denotes the $(1 - \delta)$ quantile of the standard normal distribution with CDF $F$:
$F(z_{\delta}) = 1 - \delta$. For a $95 \%$ confidence interval, $\delta = .05$ and $z_{\delta / 2} \approx 1.96 $.
In the case of small values of $K$ normal quantile should be replaced with
the one from Student's $t$ distribution on $K-1$ degrees of freedom.

Assuming that $f$ is of bounded variation, the RMSE is of the order $O(1/(\sqrt{K} n^{(1-\alpha)}))$ with
$\alpha > 0$. To obtain an accurate estimate of $\bar\mu_n$
one has to take large $n$ and small $K$ to keep the cost $nK$ at acceptable level.

\begin{table}
\centering
\caption{ MC and RQMC sample standard deviations $\sigma$, RMSE errors $\varepsilon$ and confidence intervals. The total number of function evaluations $N=nK$. }
\begin{tabular}{|c| c|}
 \hline
$\sigma_{MC} = \sqrt{ \frac{1}{(N - 1)} \sum_{i=1}^{N} (f(X_i) - \hat{\mu}_N)^2  }$ & $\sigma_{RQMC} = \sqrt{ \frac{1}{(K - 1)} \sum_{k=1}^{K} (\hat\mu_n^k - \bar\mu_n)^2  }$ \\

 \hline
 $\varepsilon_{MC} = {\frac{\sigma_{MC}}{ \sqrt N}}$ & $\varepsilon_{RQMC} =  {\frac{\sigma_{RQMC}}{ \sqrt K}} $ \\

  \hline
 $\hat{\mu}_N \pm z_{\delta / 2} \varepsilon_{MC}$ & $\bar\mu_n\pm z_{\delta / 2} \varepsilon_{RQMC}$ \\
  \hline
\end{tabular}
\label{ComparionTable_MC_RQMC}
\end{table}

Owen's nested scrambling achieves maximum 
randomization of LDS while retaining multidimensional stratification \citep{owen1997scrambled}.	
Consider a $b$-ary expansion of an LDS point in base $b$
\begin{equation} \label{LDS}
    Q_i^j = \sum_{p=1}^{m}{q_{i,p}^j b^{-p}}.
\end{equation}
Here $Q_i^j$ is $j$-th dimensional component of $Q_i$, $j=1,...,d$, $i=1,...,N$,
$N=b^m$, $b \ge 2$ and coefficients $q_{i,p}^j \in \{0,1\dots,b-1\}$.
Owen's scrambled version $V_i^j$ of $Q_i^j$ is obtained by permuting the digits $q_{i,p}^j$
in the following way:
$v_{i,1}^j = \pi^j(q_{i,1}^j)$ , $v_{i,2}^j = \pi^j_{(q_{i,1}^j)} (q_{i,2}^j)$,
$v_{i,3}^j = \pi^j_{(q_{i,1}^j,q_{i,2}^j)} (q_{i,3}^j)$, and so on. All uniform random permutations
$\pi^j$ over the set of $\{0,1\dots,b-1\}$ are mutually independent but each of them depends on
previous leading digits of $Q_i^j$.
Let M be the number of digits used in the binary number representation ($M=32$ or $M=64$).
Then the permutation tree in the $d$-dimensional case would consist of $d(b^M-1)/(b-1)$ permutations.
For Sobol' LDS with $b=2$, $M=32$ and a low dimensional problem with $d=100$ scrambling
would require to store in memory $\sim 4.3$ $10^{11}$ permutations. One way to reduce
computational costs would be to do permutations for the first $k$ bits only and then generate the other bits
randomly. In this work we use a modification of Owen's scrambling with
additional permutations \cite{Atan21}.
It has reduced memory and CPU requirements.
Owen showed in \cite{owen1997scrambled}
that for sufficiently smooth functions $\varepsilon_{RQMC} \sim O(1/(n^{(3/2-\alpha)}))$.
It is $\sqrt{n}$ times higher than the best achievable rate $O(1/(n^{(1-\alpha)}))$ for the standard (non scrambled) nets. This reduction
arises from random error cancellations.

The random digital shift (DS) method is simple to implement and it does not impose extra memory requirements as
Owen's scrambling. For simplicity we present it for the Sobol' sequence.
Consider a set $d$-dimensional Sobol' points $\{Q_i\}$  in base $b=2$ Eq.(\ref{LDS}).
Generate a random vector $U \sim U[0,1]^d$ and
produce a randomised version $V_i$ of $Q_i$ with components
$v_{i,p}^j=(q_{i,p}^j \oplus u_{p}^j)$ , $i=1,...,N, j=1,...,d, p=1,...,m$.
Here $u_{p}^j$ is the $p$-th digit in the binary representation of $U^j$. Symbol
 $\oplus$ denotes the digital addition operation (a bitwise XOR operator).
We note that $K$ randomised replicas of $\{Q_i\}$ are obtained with the same set of $\{Q_i\}$ and different $U^k$.

We note that there are other types of LDS randomization which are less costly than nested Owen’s scrambling
and more efficient than DS. A survey of these methods is given in \cite{Ecuyer18}. 
They all satisfy properties a) and b) of RQMC above, but do not possess
the increased rate of convergence of Owen's scrambling.

\section{ANOVA decomposition and effective dimension}

ANOVA decomposition can be used to explain efficiency of QMC and RQMC methods in finance.
Consider an integrable function $f(x)$ defined in the unit hypercube $H^d$. It can be decomposed as

\begin{equation}
f(x) = f_0 +  \sum_{i=1}^{d} f_i(X_i) + \sum_{i=1}^{d} \sum_{i<j}^{d} f_{ij}(X_i, X_j) + ... + f_{12....d}(X_1, X_2,...., X_d).
\end{equation}
Each of the component $f_{i_1,...,i_s}(X_{i_1},....., X_{i_s})$ is a function of a unique subset of variables from $x$.
Components $f_i(X_i)$ are called first order terms, $f_{ij}(X_i, X_j)$-second order terms and so on.
Under appropriate regularity conditions, the decomposition is unique if

\begin{equation}
\int_0^1 f_{i_1,...,i_s}(X_{i_1},....., X_{i_s}) dX_{i_k} = 0, \,\, 1 \leq k \leq s.
\end{equation}
In this case terms are orthogonal with respect to integrations \cite{BianchettiKucherenkoScoleri15}.
For square integrable functions, the total variance of $f$ decomposes as

\begin{equation}
\sigma^2 = \sum_{i=1}^{d} \sigma_i^2 + \sum_{i=1}^{d} \sum_{i<j}^{d} \sigma^2_{ij} +....+ \sigma^2_{12...d}.
\end{equation}
Here $\sigma^2_{i_1,...,i_s} = \int_0^1 f^2_{i_1,...,i_s}(X_{i_1},....., X_{i_s}) dX_{i_1}...dX_{i_s}$ are called partial variances.

Let $|u|$ be a cardinality of a set of variables $u$. Define Sobol' indices as $S_u = \frac{\sigma^2_u}{\sigma^2}$. The {\it{effective dimension}} of $f(x)$ in the
{\it{superposition sense}} is the smallest integer $d_s$ s.t $\sum_{0 < |u| < d_s} S_u \geq 1 - \epsilon$ with small $\epsilon \geq 0$.
If $d_s$ is close to 1, it means that $f$ is well approximated by a sum of $d_s$ (or less) dimensional functions.
There are cases where the first few inputs are much more important
than the others. If $\sum_{u  \subseteq {1,....,d_t} } S_u \geq 1 - \epsilon$, then $f$ has an effective dimension $d_t$
in the {\it{truncation sense}}.
Low effective dimension in the truncation
sense can sometimes be achieved by redesigning the sampling scheme in such a way that the
first few ANOVA components account for most of the variance in $f$ (see Section \ref{sec:Discretization WP} for details).

\section{Time-homogeneous hyperbolic local volatility model}

It is well known that implied volatility (the volatility input to the Black-Scholes formula that generates the market European Call or Put price)
in general depends on the strike $K$ and the maturity of the option $T$.
When implied volatility is plotted against strike price, the resulting graph is typically downward sloping for equity markets,
and the term {\it{"volatility skew"}} is often used.  For other markets, such as FX options or equity index options, where the typical graph turns up at either end, the more familiar term {\it{"volatility smile"}} is used (for details see e.g \cite{gatheral2011volatility}).
For our numerical analysis, we consider the time homogeneous hyperbolic local volatility model (HLV), which better captures the market skew. It corresponds to a parametric local volatility-type model in which the dynamic of the underlying
under the risk neutral measure $\Q$ is:

\begin{align} \label{HyperbolicSDE1}
d S(t) = rS(t) dt + \tilde{\sigma} (S(t)) dW(t), \ S_0=1,
\end{align}
where $r$ is the risk free interest rate and
\begin{equation}
\tilde{\sigma} (S) = \nu \Big\{ \frac{(1-\beta+\beta^2)}{\beta} S +\frac{(\beta-1)}{\beta}  \big(\sqrt{S^2+\beta^2(1-S)^2}-\beta\big) \Big\}.
\end{equation}
Here $\nu >0$ is the level of volatility, $\beta \in (0,1]$ is the skew
parameter and $W$ is the standard Brownian motion. This model which was introduced in \cite{Jac08} corresponds to the Black-Scholes model for $\beta=1$ and exhibits a skew for the implied volatility surface when $\beta \neq 1$.
We note that the skew increases significantly with decreasing value of $\beta$.
For example with $\nu = 0.3, \, \beta = 0.2$, the difference in volatility between strikes at $50\%$ and at $100\%$ is about $15 \%$.\\

\section{Time discretization schemes}

\subsection{Euler discretization of the SDE} \label{subsection:euler discretization}

We consider the pricing of option on a single asset whose value $S(t)$ is defined by SDE (\ref{HyperbolicSDE1}).
To guarantee positive price in the simulation,
the following transformation is used $Y(t) = \ln (S(t))$, then from (\ref{HyperbolicSDE1}) we obtain

\begin{align} \label{HyperbolicSDEY}
d Y(t) = [r -\frac{1}{2} \sigma^2(Y(t))] dt + \sigma(Y(t))     dW_t, \ Y(0)= \log (S(0)),
\end{align}
where $\sigma(Y) = \frac{\tilde{\sigma}(e^Y)}{e^Y}$.\\

For a general MC pricing framework with SDE discretization, we use Euler-Maruyama scheme
\citep{glasserman2004monte,Kloeden:Platen91}.
In a discrete case of $d$ equally distributed time steps,
it has the following form:

\begin{align} \label{HyperbolicSDEXEuler}
Y^{d}(t_{i+1}) = Y^{d}(t_{i}) +  [r-\frac{1}{2} \sigma^2(Y^{d}(t_{i}))] (t_{i+1} - t_{i}) +
\sigma(Y^{d}(t_{i})) \sqrt{t_{i+1} - t_{i}}(W(t_{i+1}) - W(t_{i}))
\end{align}
with $Y^{d}(0) = \log (S(0))$, $ \Delta t = \frac{T}{d}, t_i = i \Delta t, \, i=0,..,d$.\\

In addition to the statistical noise, there is a discretisation error.
Theorem 10.2.2 in \cite{Kloeden:Platen91}
provides conditions for Euler-Maruyama scheme to have a strong error convergence of order $\frac{1}{2}$.
Under stronger conditions as in \cite{Kloeden:Platen91},
theorem 14.5.2, the scheme reaches a weak error convergence of the order 1.

\subsection{Discretization of the Wiener process}\label{sec:Discretization WP}

There are different algorithms for the discretization of the Brownian motion $W$ in equation (\ref{HyperbolicSDEY}). The standard (incremental) discretization algorithm follows directly from the definition of $W(t)$. It is defined by the relation:

\begin{equation}\label{StdDis}
W(t_i) = W(t_{i-1}) + \sqrt{\Delta t} Z_i \,\,\, 1 \leq i \leq d,
\end{equation}
where $(Z_i)$ are independent standard normal variates obtained from random numbers or Sobol' LDS using
the inverse normal cumulative distribution function.

The Brownian bridge (BB) discretization is based on conditional distributions:
the value of $W(t_i)$ is generated from values of $W(t_l),W(t_m), l \leq i \leq m$ at earlier and later time steps.
This discretization first generates the Brownian motion at the terminal point
$$W(T) = \sqrt{T}Z_1$$
and then it fills other points using already found values of $W(t_i)$. The generalised BB formula is given by
\begin{equation} \label{BBFormula}
W(t_i) = (1 - \gamma)W(t_l) + \gamma W(t_m) + \sqrt{\gamma(1 - \gamma)(m - l) \Delta t}Z_i,
\end{equation}
where $\gamma = \frac{i-l}{m-l}$. It can be seen from equation (\ref{BBFormula}) that
the variance of the stochastic part of the BB formula
$\gamma(1 - \gamma)(m - l) \Delta t$ decreases rapidly at the successive levels of refinement
and the first few points contain most of the variance.
Moreover, the variance in the stochastic part of (\ref{BBFormula}) is smaller than that in
(\ref{StdDis}) for the same time steps.

For MC the BB scheme has the same efficiency as the standard one but it does affect the efficiency of
QMC based on Sobol' LDS. 
Sobol’ defined “Sobol’ sequence” as the $LP\tau$ sequence. The $\tau$-value is a quality parameter
which measures the uniformity of the point sets. 
The smaller the $\tau$-value is the more uniformly distributed the points are. 
This value is equal to 0 only for one and two dimensional Sobol’ sequences. 
In higher dimensions, as $d$ increases, the smallest possible values of $\tau$ increase as well.
Hence, the initial coordinates of Sobol' LDS are much better distributed than the
later high dimensional coordinates.

The BB discretization uses
low well distributed coordinates from each $d$-dimensional LDS vector point to determine
most of the structure of a path and reserves the later coordinates to fill
in fine details. In other words, well distributed coordinates are used for
important variables and higher not so well distributed coordinates are
used for far less important variables.
Thus the BB sampling reduces the effective dimension in the truncation sense which leads
to the much higher convergence rate of the QMC algorithm for majority (but not all) of the payoffs
\citep{BianchettiKucherenkoScoleri15}.

\section{Monte Carlo simulation of option pricing and Greeks}

%

\subsubsection{Option pricing} \label{subsect:option_pricing}

We consider a geometric average Asian call option whose payoff function is given by

\begin{equation} \label{asianpayoff}
P_A = \max (\bar{S} - K, 0),
\end{equation}
where $\bar{S}$ is a geometric average at $d$ equally spaced time point:

\begin{equation} \label{geometricaverage}
\bar{S} = ( \prod_{i=1}^d S_i)^{\frac{1}{d}},
\end{equation}
where $S_i$ is the asset price at time $t_i = i \frac{T}{d}$, $1 \leq i \leq d$.

In a risk neutral setting, the value of a call option with maturity $T$ and strike $K$ is the discounted
value of its payoff:

\begin{equation} \label{priceasianpayoff}
AC(T,K) = e^{-rT} \E^{\Q}[P_A].
\end{equation}

There is no analytical formula for (\ref{priceasianpayoff}) in the HLV model and it is
estimated by the MC method. Firstly, we approximate the asset price $S(t_i)$ with $S^d(t_i) = e^{Y^d(t_i)}$ by discretising
the SDE (\ref{HyperbolicSDEY}) as described in Section \ref{subsection:euler discretization}.
Secondly, the expectation of the Asian payoff (\ref{asianpayoff}) is computed with the MC estimator as
an arithmetic average of payoffs taken over a finite number $N$ of simulated price path:

\begin{equation} \label{MCpriceasianpayoff}
AC_N(T,K) = e^{-rT} \left[ \frac{1}{N} \sum_{i=1}^N \max (\bar{S}^{(i)} - K, 0) \right],
\end{equation}
where $\bar{S}^{(i)}$ is an approximation of $\bar{S}$ using the simulated $i$-th price paths.

\subsubsection{Sensitivity factors}

Sensitivity factors or {\it{Greeks}} are derivatives of the price $AC(T,K)$ w.r.t specific parameters like
spot price or volatility. They are computed for hedging and risk management purposes.
In this work, we focus only on the Delta defined as
$\Delta = \frac{\partial AC(T,K) }{\partial S_0}$, where $S_0$ is the current spot price.
In the dynamic hedging, Delta corresponds to the number of assets one should hold
for each option shorted to maintain a delta-neutral position. As there is no analytical formula for the
value of $AC_N(T,K)$, Delta can be estimated by MC simulation and the finite difference method.
In the case of the central difference scheme Delta is computed as

\begin{equation}
\Delta \approx \frac{ AC_N(T,K, S_0 + \epsilon_s) - AC_N(T,K, S_0 - \epsilon_s) }{2 \epsilon_s},
\end{equation}
where $\epsilon_s = h S_0 $ is the increment, $h$ is a shift parameter.

Path recycling of both pseudo-random sequences and LDS is used to
minimize the variance, as suggested e.g. in \cite{glasserman2004monte}.
We note that the error analysis for Greeks is more complex
than that for prices, since the variance of the MC simulation is mixed with the bias due
to the approximation of derivatives with finite differences.
For the sensitivity factor estimation
not to be entirely hidden by the MC noise, in our computations
the shift is chosen to be large enough: $h=0.01$ (see \cite{glasserman2004monte} for detailed discussions).

\section{Numerical results}

In this Section we present the results from simulations of
prices and sensitivity factor $\Delta$ for Asian call options on a single underlying.
The following parameters
were used in simulations: $S_0=100, \, r = 3 \%, \, T = 1, \ \nu = 30\%, \, \beta = 0.5$,
number of discrete time steps $d$ = 256. In the single underlying case,
$d$ corresponds to the problem dimensionality. We consider {\it{in-the-money}} (ITM),
{\it{at-the-money}} (ATM) and {\it{out-the-money}} (OTM)
options with strike $80, \, 100, \, \120$ respectively to investigate the effect of moneyness.

Numerical simulations using MC, QMC and RQMC methods were performed to compare
convergence of each method. The standard (incremental) discretisation of Brownian motion was used in the MC method.
The Brownian Bridge algorithm was used in QMC and RQMC methods (Section \ref{sec:Discretization WP}).
The Mersenne Twister generator was used for MC simulations and RQMC with digital shift.
BRODA’s Sobol' sequence generator with additional uniformity properties described in Section \ref{sec:MC, QMC and RQMC methods}
was used for QMC simulations \cite{BRODA,SobAsoKreiKuch11}.
For QMC and RQMC to achieve an optimal uniformity sampling, the number of points
$n$ was taken to be powers of two. The reference values of prices and Deltas were obtained by the MC method
by averaging over $K = 10$ independent runs with each run using $n=2^{18}$ paths.

\begin{figure}
    \centering
       \subfigure[ITM]{\includegraphics[scale=0.20,trim={0cm 0.0cm 0cm 0cm}, clip]{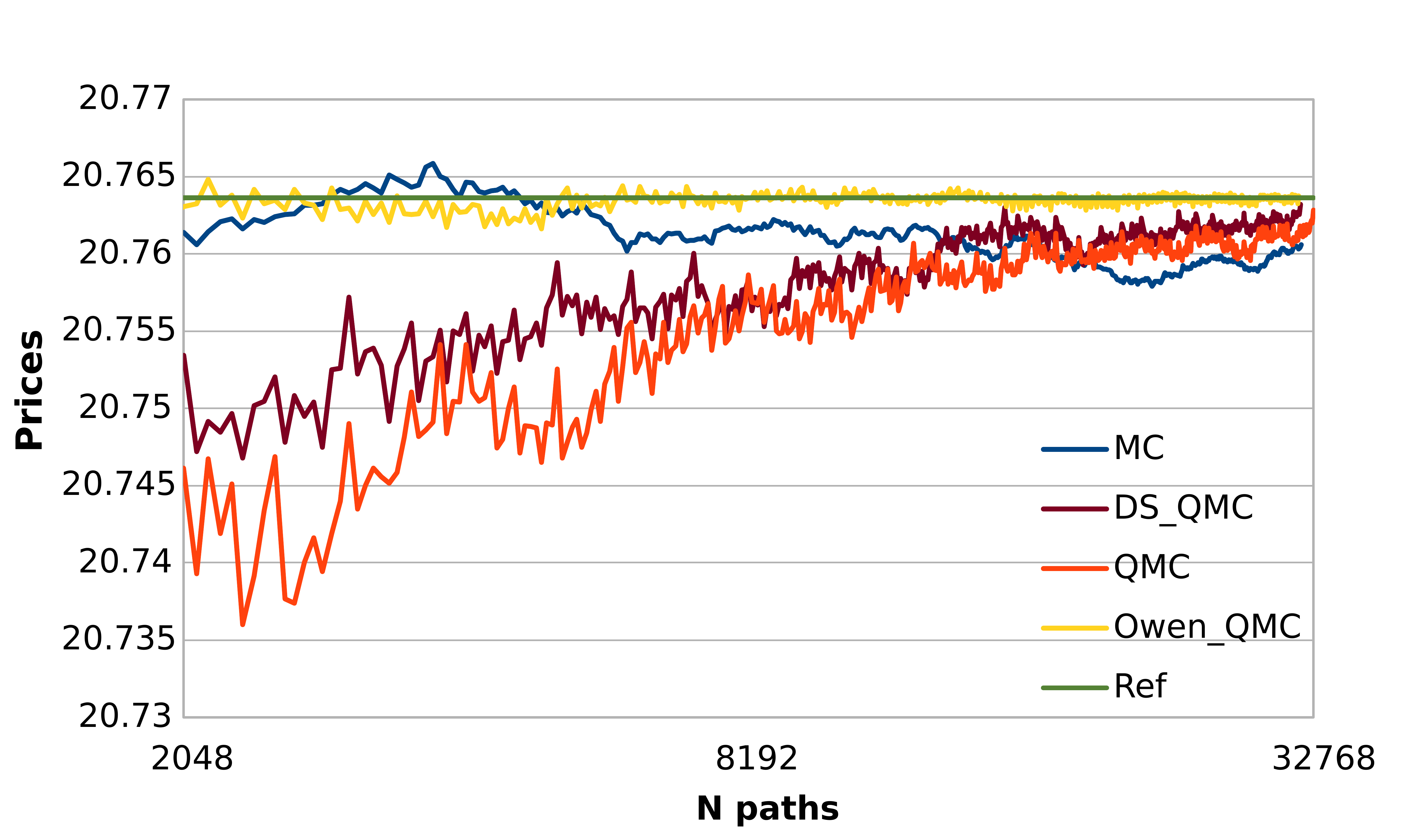}}
       \subfigure[ATM]{\includegraphics[scale=0.20,trim={0cm 0.0cm 0cm 0cm}, clip]{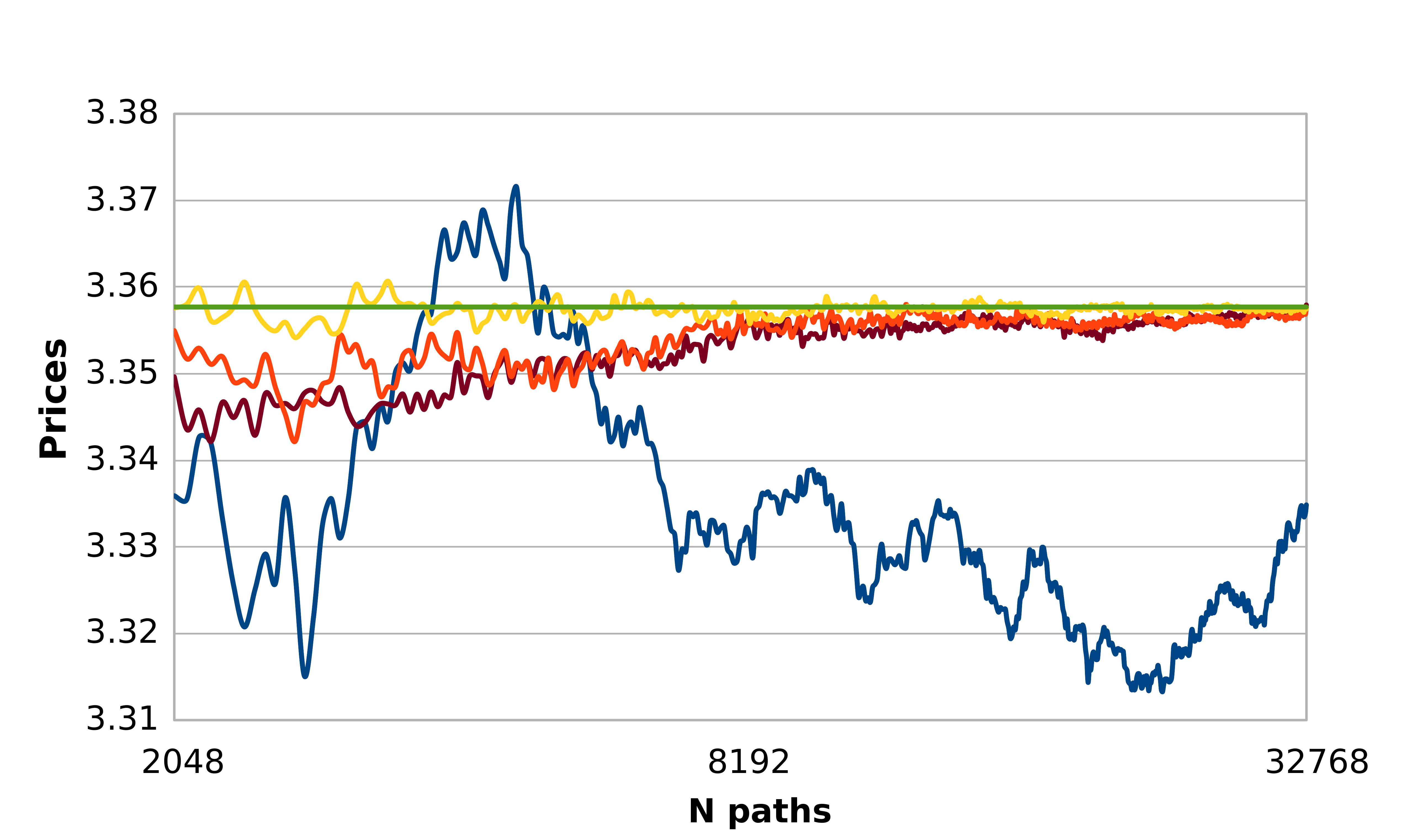}}
       \subfigure[OTM]{\includegraphics[scale=0.20,trim={0cm 0.0cm 0cm 0cm}, clip]{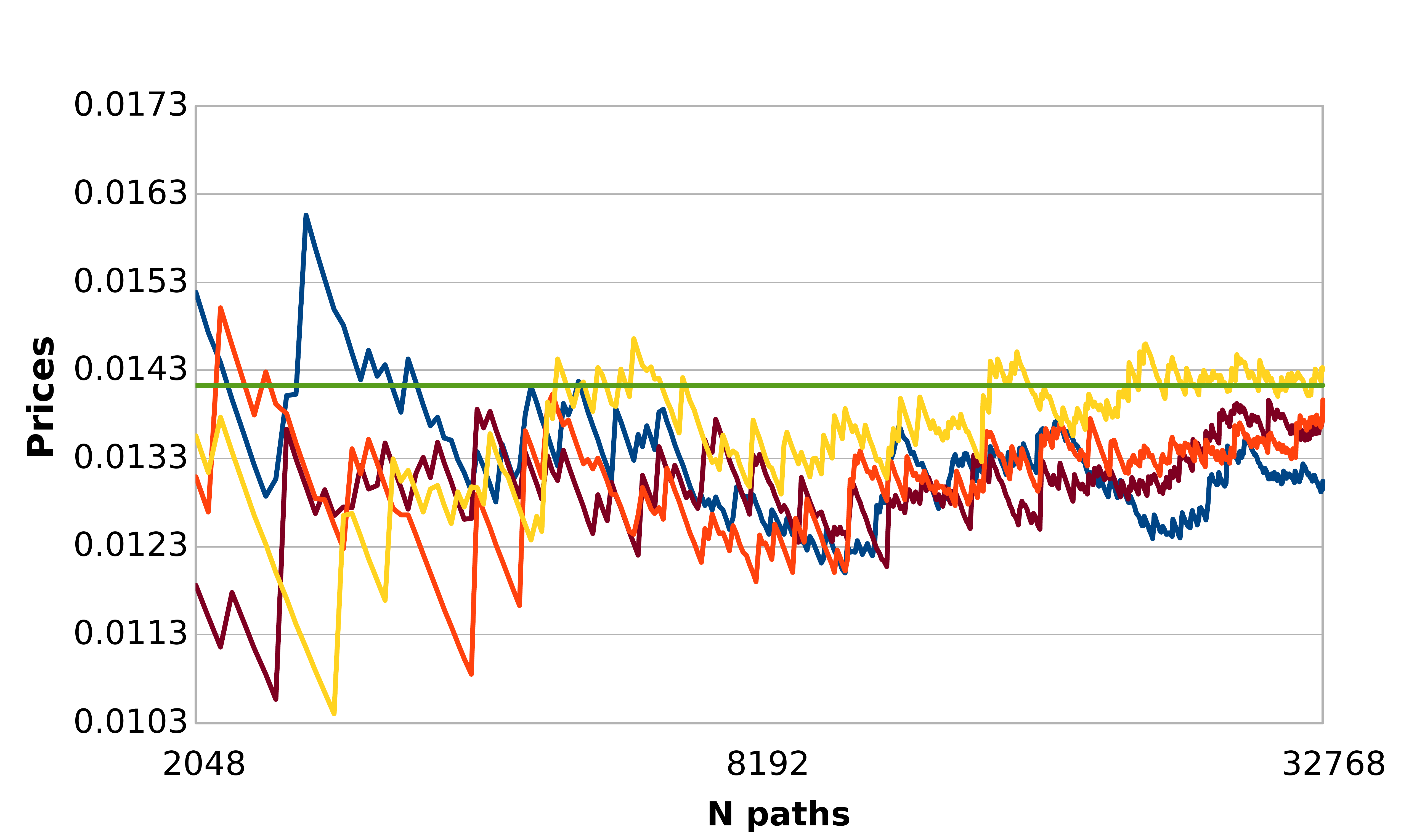}}
    \caption{Asian call price $(a)$ ITM $(b)$ ATM $(c)$ OTM w.r.t number of paths $N=n$, $K=1$ (in $log2$ scale).}
    \label{Fig_ITMAsianCallprices_Zoom}
\end{figure}

\begin{figure}
    \centering
        \subfigure[ITM]{\includegraphics[scale=0.20,trim={0cm 0.0cm 0cm 0cm}, clip]{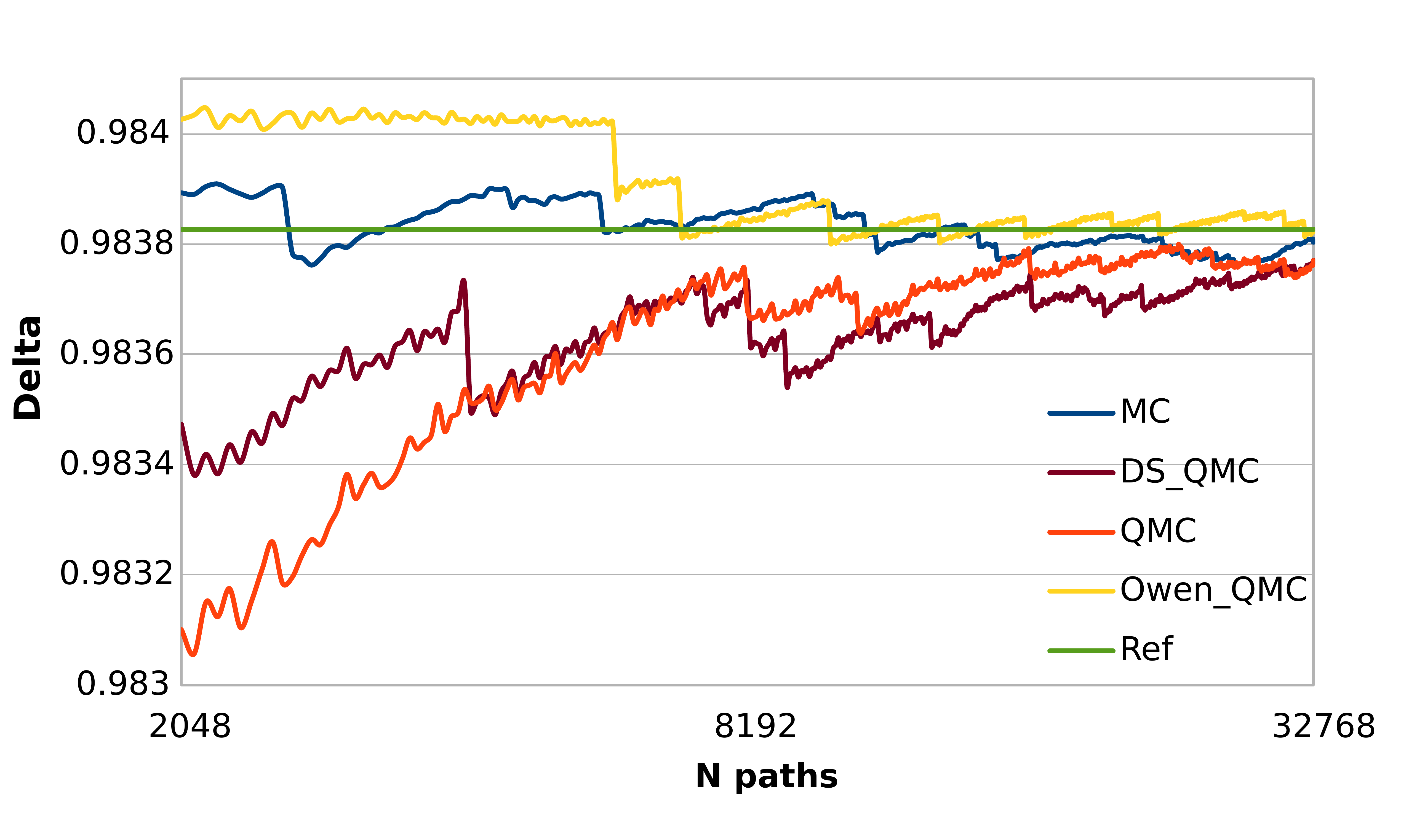}}
        \subfigure[ATM]{\includegraphics[scale=0.20,trim={0cm 0.0cm 0cm 0cm}, clip]{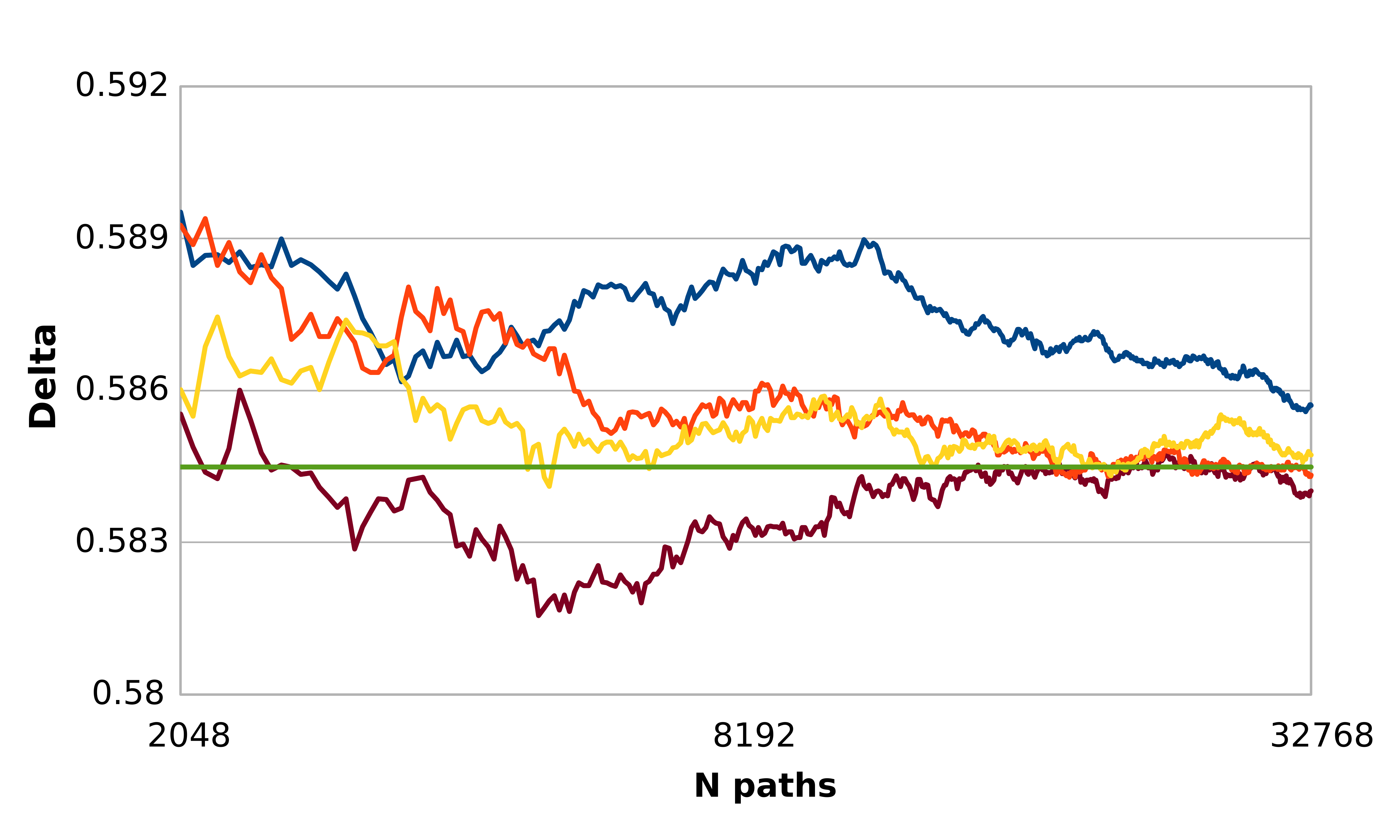}}
        \subfigure[OTM]{\includegraphics[scale=0.20,trim={0cm 0.0cm 0cm 0cm}, clip]{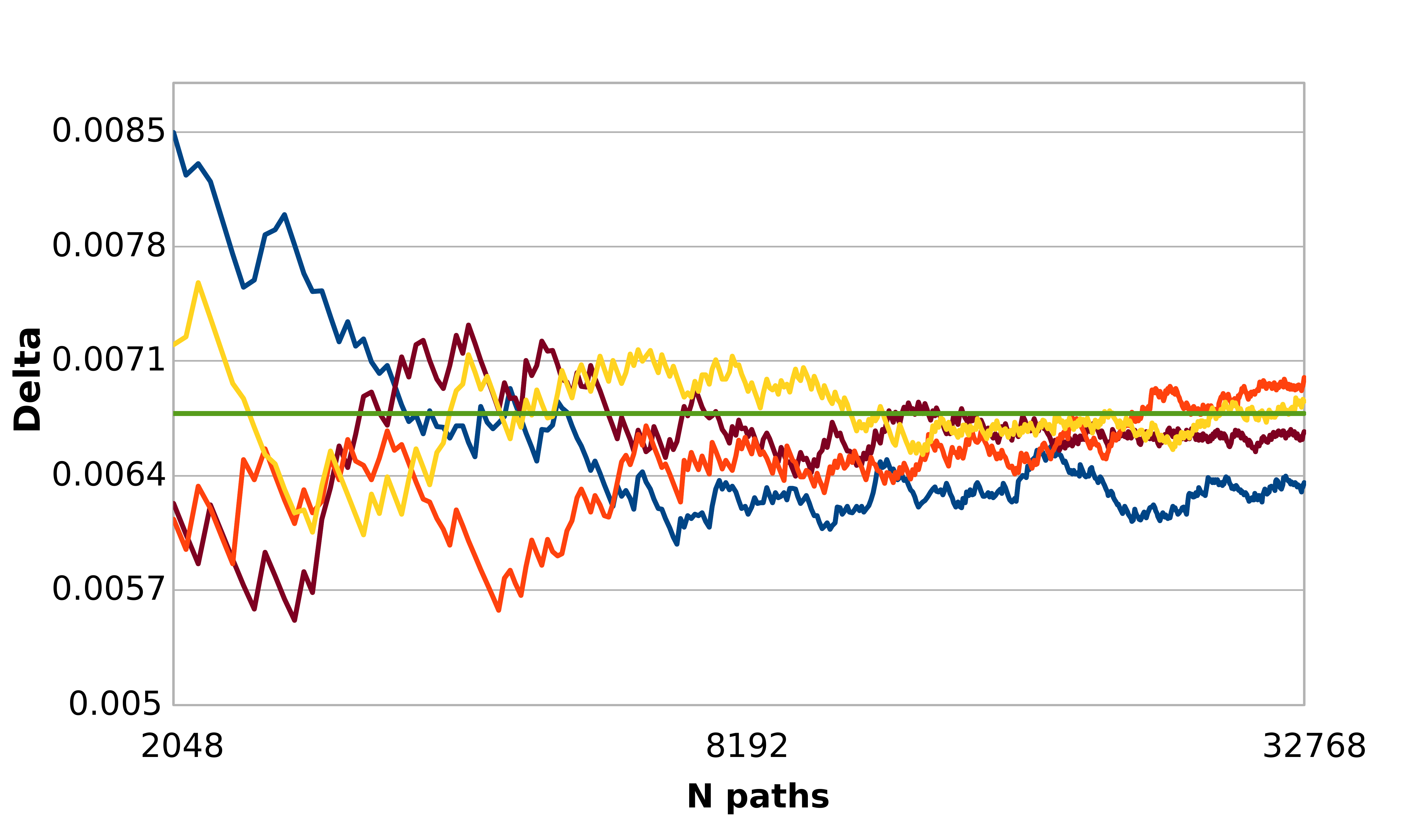}}
    \caption{Asian call Delta $(a)$ ITM $(b)$ ATM $(c)$ OTM w.r.t number of paths $N=n$, $K=1$ (in $log2$ scale).}
    \label{Fig_ITMAsianCallDelta_Zoom}
\end{figure}

\subsection{Price and delta convergence}

Firstly, we analyze convergence plots, namely values of price (Figs. \ref{Fig_ITMAsianCallprices_Zoom})
and Delta (Figs. \ref{Fig_ITMAsianCallDelta_Zoom}) versus the number of paths. One trial ($K=1$) is used for MC and RQMC runs. For simulated prices RQMC with Owen's scrambling outperforms all other methods for the ITM and ATM cases
converging quicker to the reference levels. Its efficiency is followed by QMC and RQMS with DS methods, which outperform
MC for the ITM and ATM cases. These two methods show similar performance between themselves.
Similar but less pronounced trends are present in the case of OTM.

For simulated Deltas RQMC with Owen's scrambling marginally outperform all other methods for the ITM case.
It shows a similar performance to QMC for the ATM case. Both RQMC and QMC methods slightly
outperform MC for the case of OTM.

We note, that the step-like behaviour of the convergence patterns of in some QMC, RQMC plots  
is likely to be a result of the inherent design of Sobol’ sequence generators as can be seen from Fig. 2, p. 70
in \cite{SobAsoKreiKuch11}.

We can conclude that all considered  cases (ITM, ATM and OTM), RQMC method with Owen's scrambling
shows a faster convergence than other methods. RQMC with DS shows similar to QMC performance.

\subsection{Confidence intervals}

\begin{table}
\centering
\caption{ $\varepsilon_{MC}$ and $\varepsilon_{RQMC}$ at $K$=10, $n$=$2^{12}$ ($N$=40960) of price estimations.}
\begin{tabular}{|l|c|c|c|}
\hline 
& ITM& ATM & OTM \\
\hline
$\varepsilon_{MC}$ & 3.26 $10^{-2}$  & 2.2 $10^{-2}$ & 1.24 $10^{-3}$ \\
 \hline
$\varepsilon_{RQMC}$ (Owen) & 3.67 $10^{-4}$  & 6.09 $10^{-4} $  & 4.04 $10^{-4}$ \\
\hline
$\varepsilon_{RQMC}$ (DS)  & 7.12 $10^{-4}$  & 1.03 $10^{-3} $  & 4.59 $10^{-4}$ \\
\hline
$\varepsilon_{MC} / \varepsilon_{RQMC}$ (Owen) & 89 & 36 & 3 \\
\hline
\end{tabular}
\label{tab:average_sample standard deviation}
\end{table}

Tables \ref{tab:average_sample standard deviation}, \ref{tab:average_sample standard deviation for delta}
show RMSE (Table \ref{ComparionTable_MC_RQMC}) for MC and RQMC methods of prices and Deltas estimations, respectively.
Ratios  of MC to RQMC with Owen's scrambling error estimates show a dramatic improvement with using RQMC with the largest
improvement ratio for ITM. RQMC with Owen's scrambling method on average is better than RQMC with DS producing smaller $\varepsilon_{RQMC}$.

\begin{table}
\centering
\caption{ $\varepsilon_{MC}$ and $\varepsilon_{RQMC}$ at $K$=10, $n$=$2^{12}$ ($N$=40960) of Deltas estimations.}
\begin{tabular}{|l|c|c|c|}
\hline 
& ITM& ATM & OTM \\
\hline
$\varepsilon_{MC}$ & 3.29 $10^{-4}$  & 2.43 $10^{-3}$ & 4.09 $10^{-4}$ \\
 \hline
$\varepsilon_{RQMC}$ (Owen) & 3.23 $10^{-5}$  & 5.98 $10^{-4} $  & 1.41 $10^{-4}$ \\
\hline
$\varepsilon_{RQMC}$ (DS)  & 4.04 $10^{-5}$  & 4.34 $10^{-4} $  & 1.46 $10^{-4}$ \\
\hline
$\varepsilon_{MC} / \varepsilon_{RQMC}$ (Owen) & 10 & 4 & 3 \\
\hline
\end{tabular}
\label{tab:average_sample standard deviation for delta}
\end{table}

\begin{figure}
    \centering
       \subfigure[RQMC Owen]{\includegraphics[scale=0.25,trim={0cm 0.2cm 0cm 0cm}, clip]{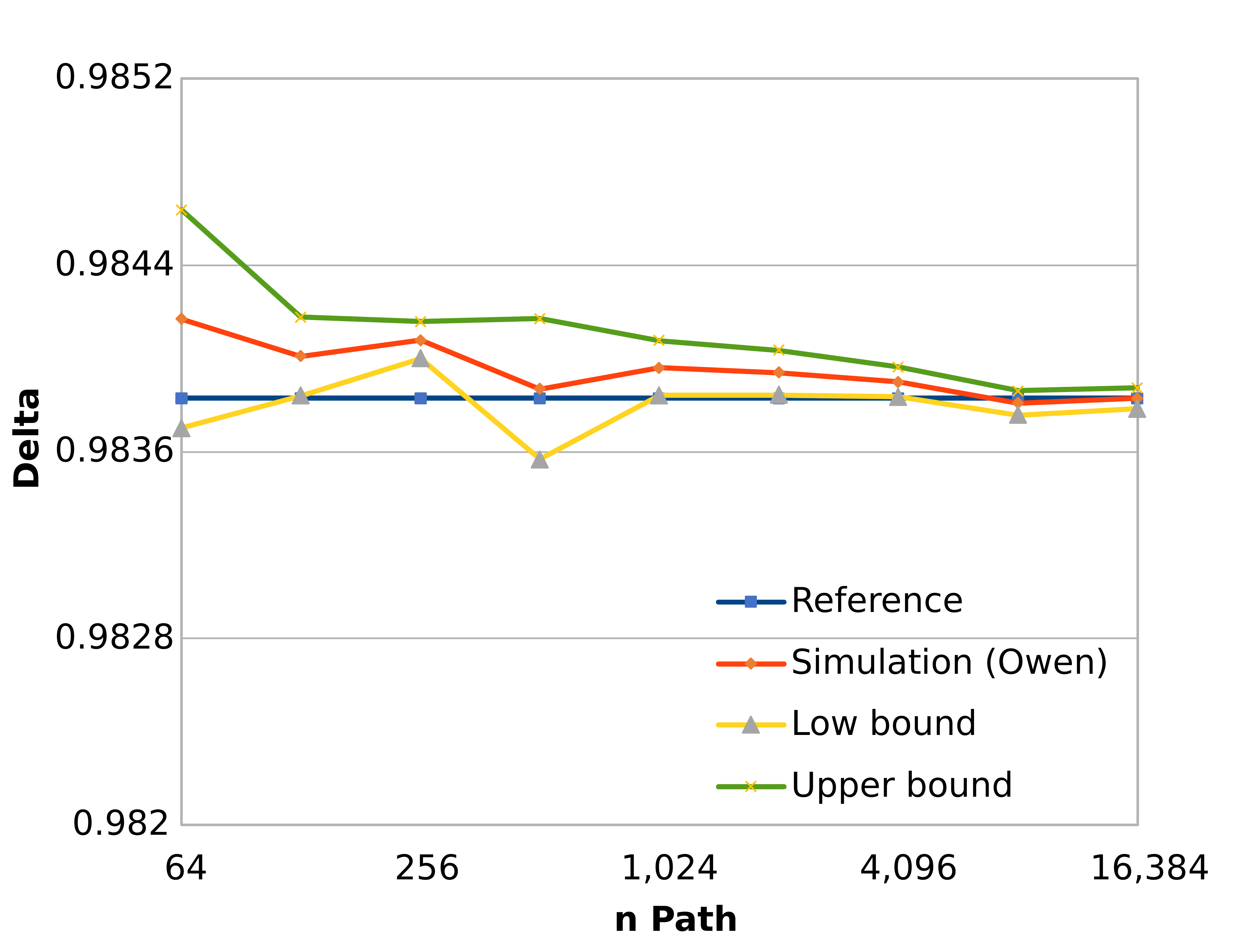}}
       \subfigure[RQMC DS]{\includegraphics[scale=0.25,trim={0cm 0.2cm 0cm 0cm}, clip]{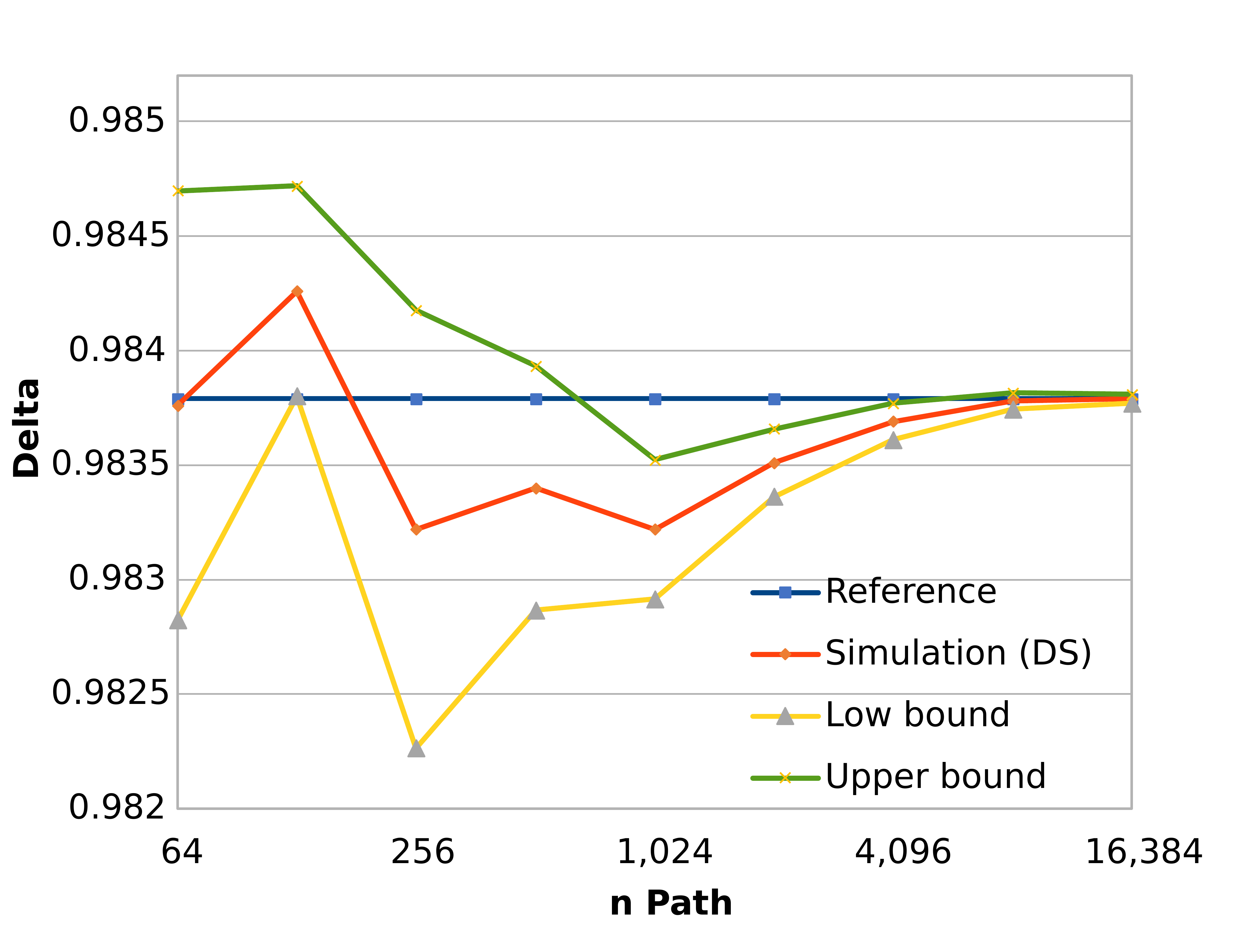}}
    \caption{Asian Call Delta with 95\% confidence intervals computed  at $K=10$ for
    ITM call for RQMC $(a)$ Owen $(b)$ DS methods w.r.t number of paths $n$ (in $log2$ scale).}
    \label{Fig_MC_CI_ITMAsianCall}
\end{figure}

Prices and Deltas with confidence intervals versus the number of simulation paths $n$ for the ITM call 
are shown in Figures \ref{Fig_MC_CI_ITMAsianCall}.
Visually the results for the ATM and OTM cases are similar and they are not shown.
As expected the confidence intervals are tightening with increasing the number of paths.
RQMC with Owen's scrambling offers the most accurate results by reducing significantly the bounds of confidence intervals.

\subsection{Performance analysis}

We also analyze the relative performance of considered methods in terms of convergence rates.
For all considered sampling schemes the following power law for the integration error is observed
empirically in numerical tests:
\begin{equation}
\label{err:QMC}
\varepsilon_n \sim \frac{C}{n^\alpha}\ .
\end{equation}
For the MC method $\alpha = 0.5$. For applications of the QMC and RQMC methods to financial problems quite commonly
$\alpha>0.5$. Its value can be very close to 1
irrespective of the nominal dimension when the effective dimensions are low.

It has been discussed in Section \ref{sec:MC, QMC and RQMC methods} that
there are no statistical measures like variances associated with LDS because they are deterministic.
Hence, the constant $C$ in (\ref{err:QMC}) is not a variance and
(\ref{err:QMC}) does not have a probabilistic interpretation.
In practice, the root mean square error (RMSE) for both MC, RQMC and QMC methods for
any fixed $n$ can be estimated by computing the following error
averaged over $K$ independent runs:
\begin{equation}\label{error}
\varepsilon_n=\sqrt{\frac{1}{K}\,\sum_{k = 1}^K\left(V-V_n^{(k)}\right)^2},
\end{equation}
where $V$ is the exact, or estimated value of the integral (option price or Delta in our case)
obtained at a very large $n\rightarrow\infty$, $V_n^{(k)}$ is the simulated
value for the $k$-th run, performed using $n$ paths.

We note some difference between definitions of $\varepsilon_{RQMC}$ 
given in Table 1 (it is computed with a reference to 
$\bar\mu_n$) and $\varepsilon_n$  in (\ref{error}) (it is computed with a reference to $V$). 

For MC and RQMC, runs based on different seed points are statistically
independent. In the case of QMC, different runs are obtained using
non overlapping sections of the LDS. In our computations $K$ = 10.

\begin{table}
\centering
\caption{Extracted $\alpha$ in QMC and RQMC (Owen's scrambling) methods.}
\begin{tabular}{|l|c|c|c|}
\hline 
& ITM& ATM & OTM \\
\hline
QMC (Price) & 1.0 & 0.95 & 0.82\\
\hline
RQMC (Price) & 0.7 & 0.79 & 0.79 \\
\hline
QMC (Delta) & 0.65 & 0.64 & 0.71\\
\hline
RQMC (Delta) & 0.66 & 0.62 & 0.62 \\
\hline
\end{tabular}
\label{tab:alfa_conv}
\end{table}

\begin{figure}
    \centering
       \subfigure[ITM]{\includegraphics[scale=0.22,trim={0cm 0.0cm 0cm 0cm}, clip]{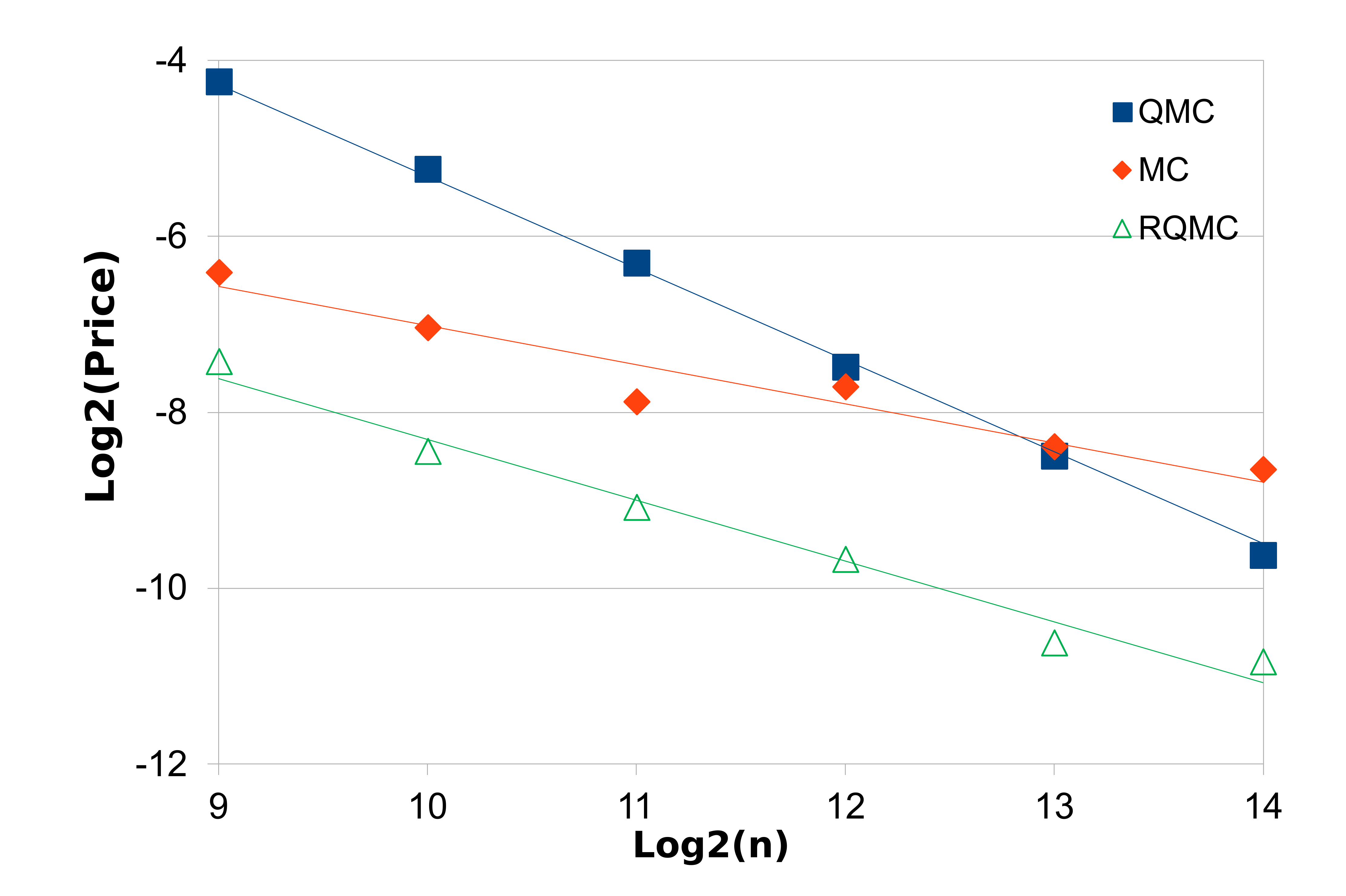}}
       \subfigure[ATM]{\includegraphics[scale=0.22,trim={0cm 0.0cm 0cm 0cm}, clip]{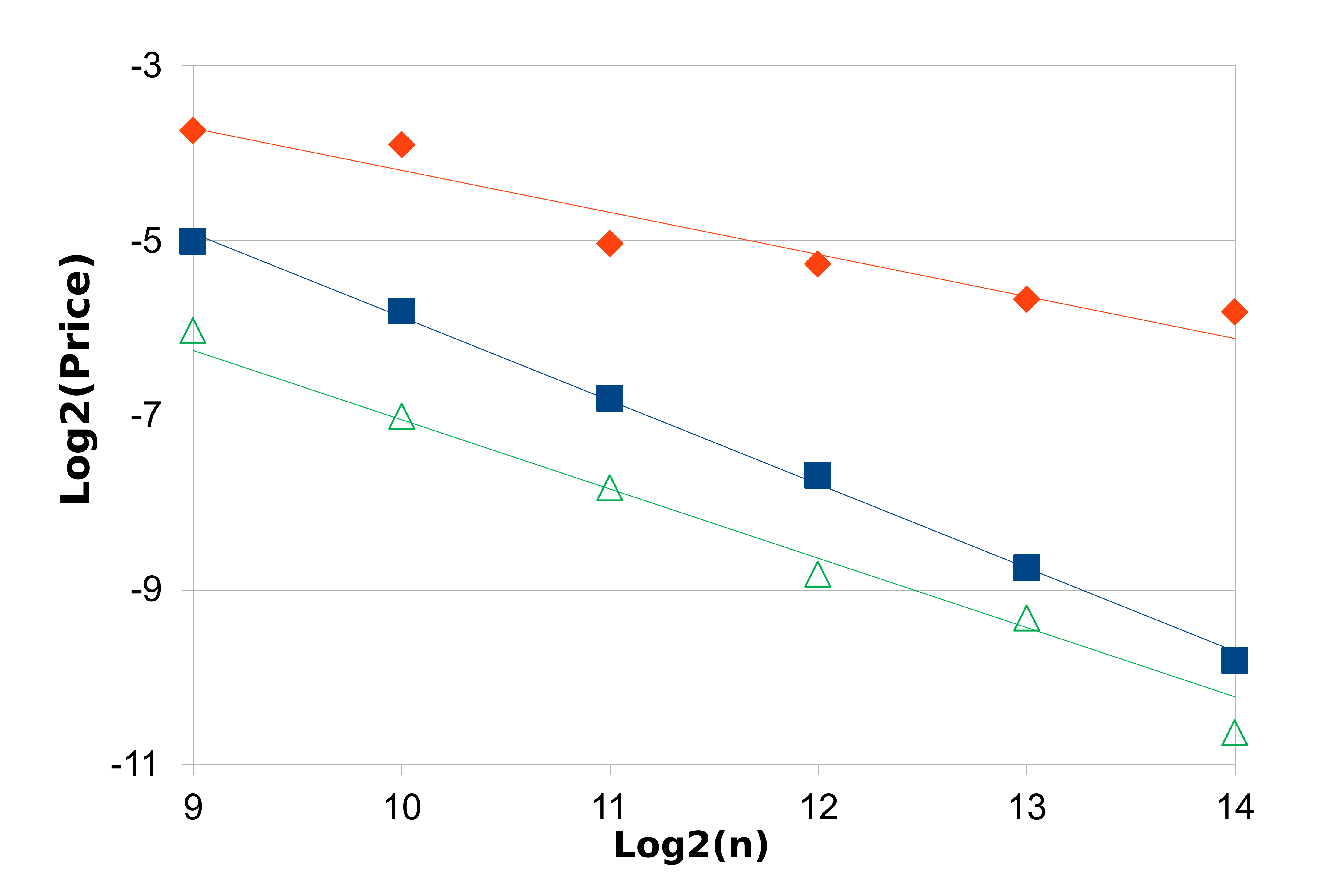}}
       \subfigure[OTM]{\includegraphics[scale=0.22,trim={0cm 0.0cm 0cm 0cm}, clip]{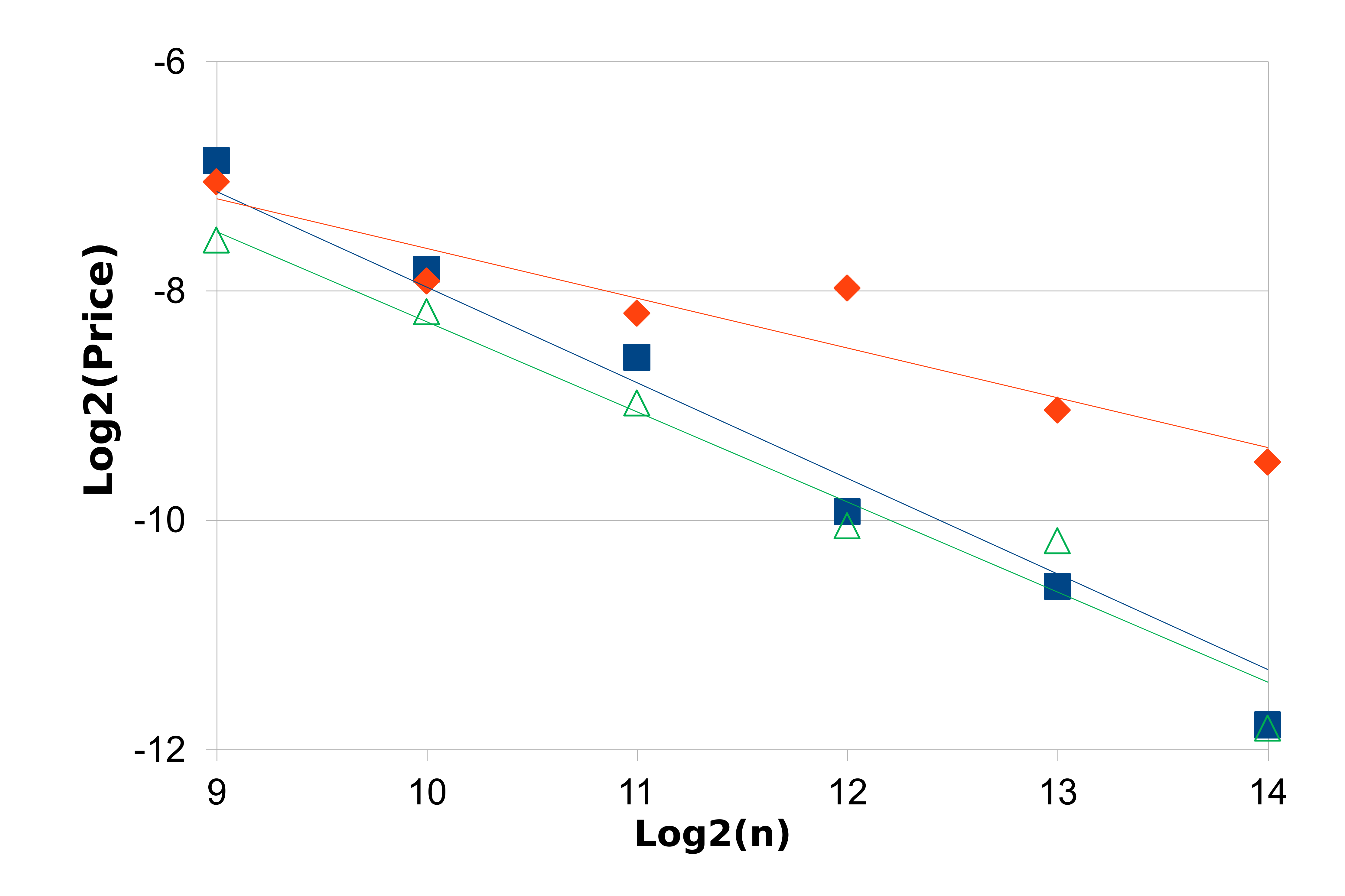}}
    \caption{RMSE for ITM Asian call Prices $(a)$ ITM  $(b)$ ATM $(c)$ OTM w.r.t number of paths $n$, $K=10$.
    Owen's scarmbling is used in RQMC.}
    \label{Fig_RMSE_Price}
\end{figure}

\begin{figure}
    \centering
       \subfigure[ITM]{\includegraphics[scale=0.22,trim={0cm 0.0cm 0cm 0cm}, clip]{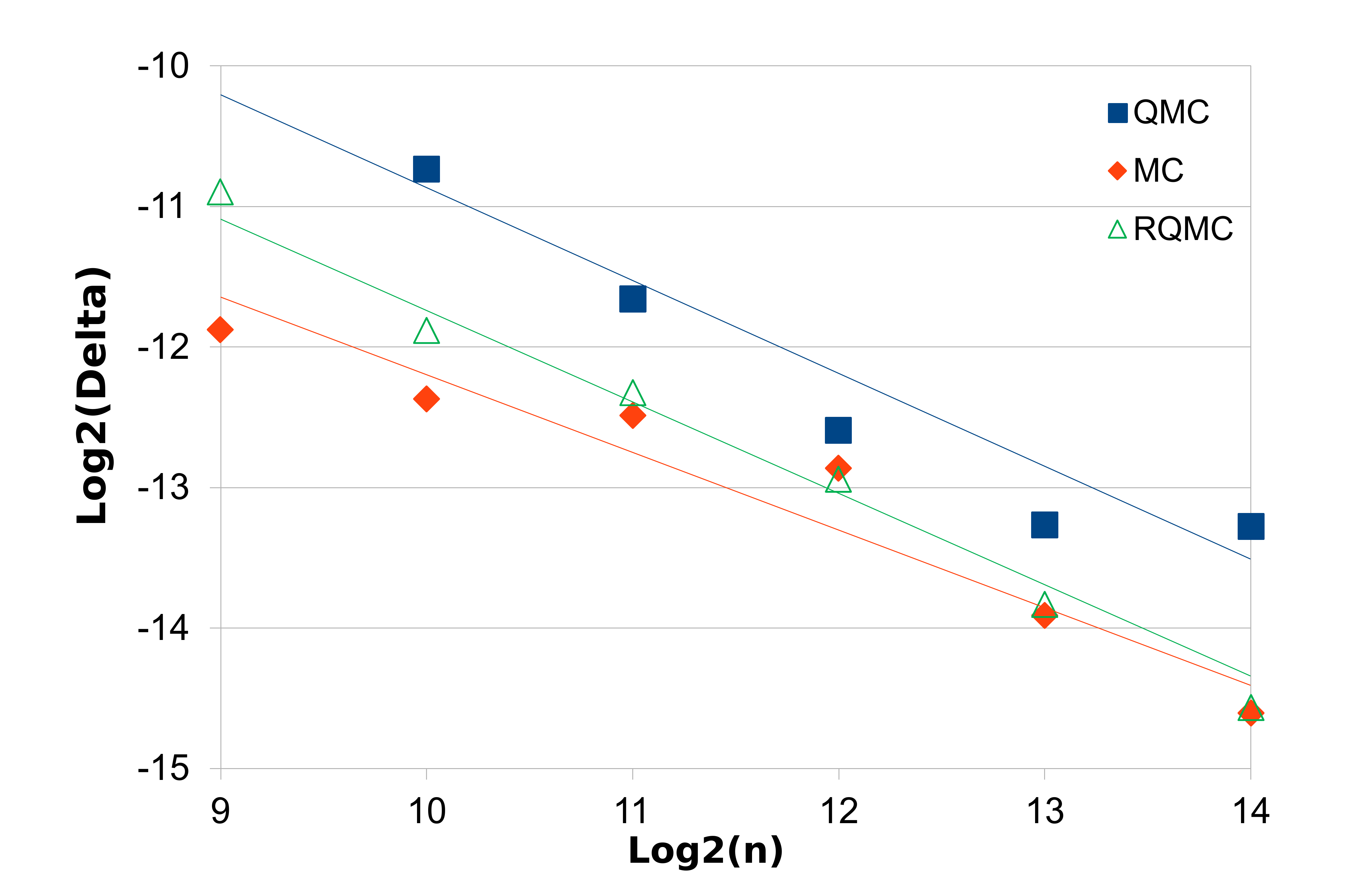}}
       \subfigure[ATM]{\includegraphics[scale=0.22,trim={0cm 0.0cm 0cm 0cm}, clip]{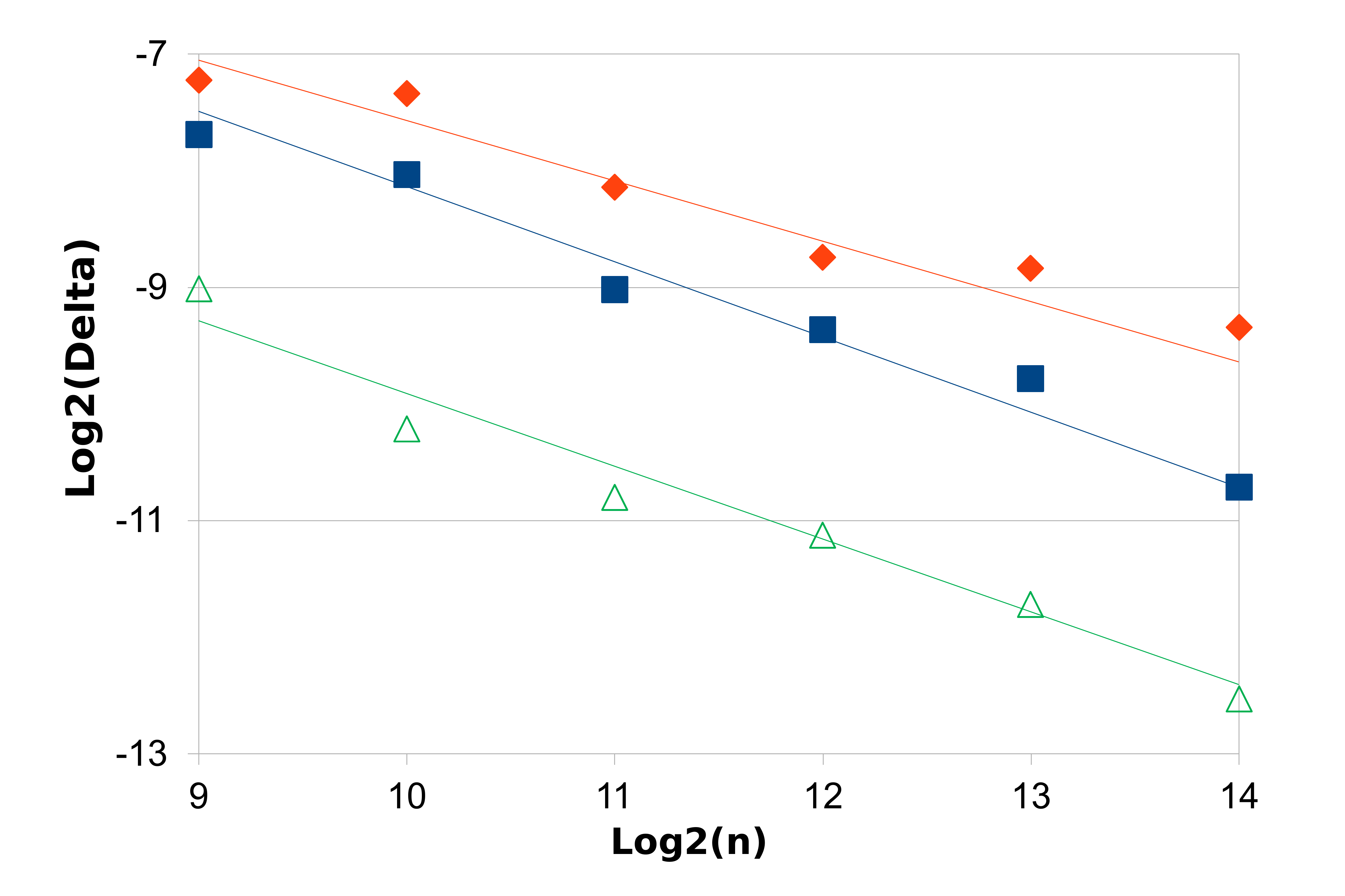}}
       \subfigure[OTM]{\includegraphics[scale=0.22,trim={0cm 0.0cm 0cm 0cm}, clip]{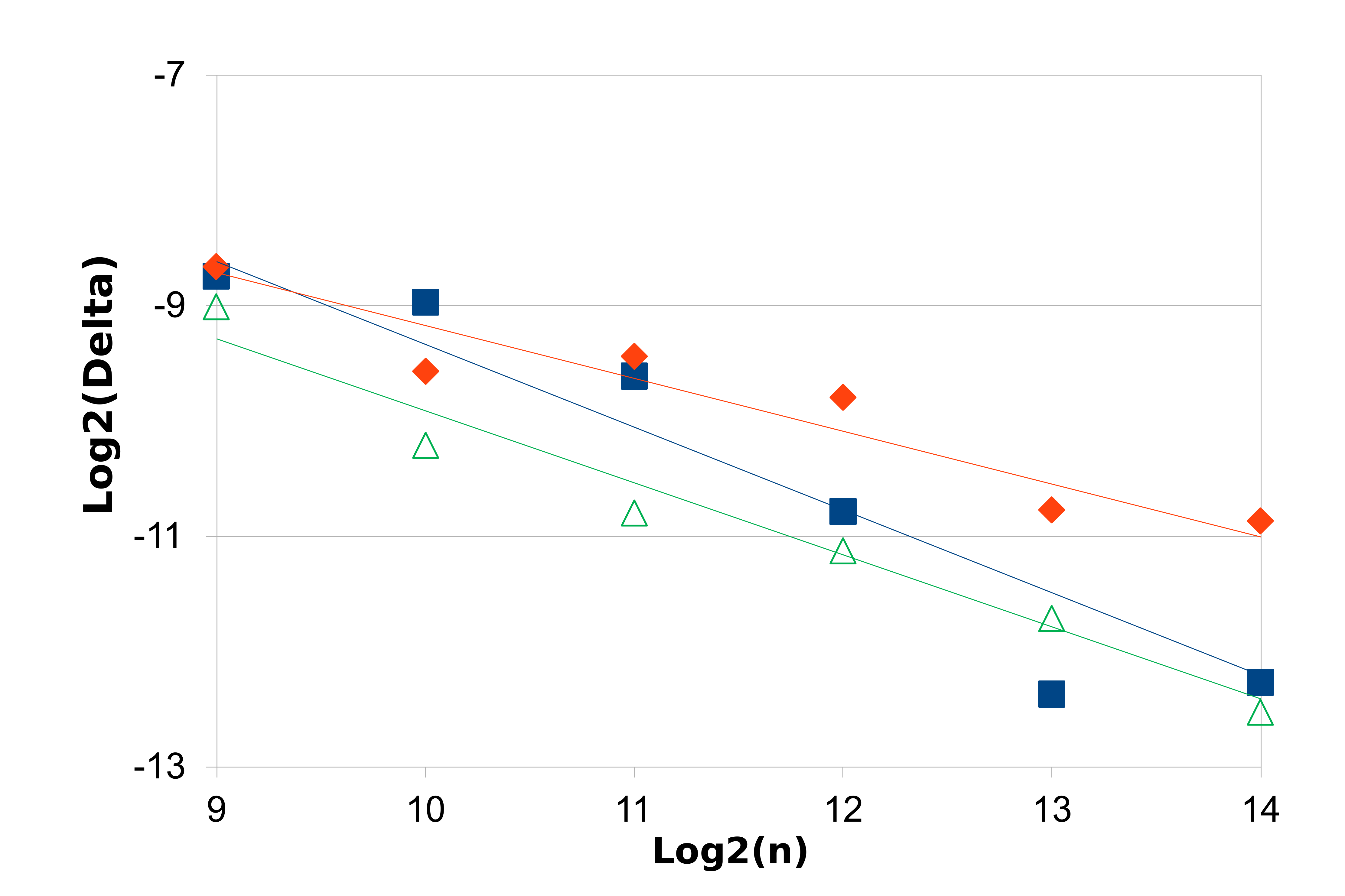}}
    \caption{RMSE for ITM Asian call Deltas $(a)$ ITM  $(b)$ ATM $(c)$ OTM w.r.t number of paths $n$, $K=10$.
    Owen's scarmbling is used in RQMC.}
    \label{Fig_RMSE_Delta}
\end{figure}

Figures \ref{Fig_RMSE_Price}, \ref{Fig_RMSE_Delta} show the RMSE versus the number of paths $n$ for MC, QMC and RQMC methods in $log2–log2$ scale. We fitted the regression lines that follow the power law
(\ref{err:QMC}) to extract convergence rates $\alpha$: they are
the slopes of the regression lines (Table \ref{tab:alfa_conv}). We also extracted the intercepts of regression lines
($log2(C)$, (\ref{err:QMC})) (not presented here). They provide useful information about the efficiency
of the QMC, RQMC and MC methods: lower intercepts mean that the simulated value
starts closer to the exact value.

As expected, for MC $\alpha=0.5$ for all cases, these results are not presented in
the Table \ref{tab:alfa_conv}. For QMC and RQMC $\alpha > 0.5$ for all cases. It is higher for price than for Delta.
Although it is marginally higher for QMC but the intercepts $C$ of regression lines
are always lower for RQMC than for other methods for considered ranges of $n$.
It makes RQMC the most efficient method (although efficiencies of RQMC and QMC
are similar for the OTM case).

\section{Conclusions}
We present and discuss the results of an application of MC, QMC and RQMC methods for derivative
pricing and risk analysis based on the hyperbolic local volatility model. The results presented for the Asian option show the superior
performance of the QMC and RQMC methods. RQMC not only increases the rate of convergence of QMC but also
allows to compute confidence intervals around the estimated value.
Efficiency of RQMC strongly depends on the scrambling methods. We advise
to use Sobol' LDS with Owen's scrambling as the most efficient method.

\newpage

\bibliographystyle{apacite}
\bibliography{bibliography}

\begin{thebibliography}{}

\bibitem [\protect \citeauthoryear {%
Atanassov%
\ \BBA {} Kucherenko%
}{%
Atanassov%
\ \BBA {} Kucherenko%
}{%
{\protect \APACyear {2021}}%
}]{%
Atan21}
\APACinsertmetastar {%
Atan21}%
\begin{APACrefauthors}%
Atanassov, E.%
\BCBT {}\ \BBA {} Kucherenko, S.%
\end{APACrefauthors}%
\unskip\
\newblock
\APACrefYearMonthDay{2021}{}{}.
\newblock
{\BBOQ}\APACrefatitle {Implementation of {O}wen's scrambling with additional
  permutations for {S}obol' sequences} {Implementation of {O}wen's scrambling
  with additional permutations for {S}obol' sequences}.{\BBCQ}
\newblock
\APACjournalVolNumPages{BRODA Ltd., UK}{}{}{}.
\PrintBackRefs{\CurrentBib}

\bibitem [\protect \citeauthoryear {%
Bianchetti%
, Kucherenko%
\BCBL {}\ \BBA {} Scoleri%
}{%
Bianchetti%
\ \protect \BOthers {.}}{%
{\protect \APACyear {2015}}%
}]{%
BianchettiKucherenkoScoleri15}
\APACinsertmetastar {%
BianchettiKucherenkoScoleri15}%
\begin{APACrefauthors}%
Bianchetti, M.%
, Kucherenko, S.%
\BCBL {}\ \BBA {} Scoleri, S.%
\end{APACrefauthors}%
\unskip\
\newblock
\APACrefYearMonthDay{2015}{}{}.
\newblock
{\BBOQ}\APACrefatitle {Pricing and Risk Management with High-Dimensional
  Quasi-Monte Carlo and Global Sensitivity Analysis} {Pricing and risk
  management with high-dimensional quasi-monte carlo and global sensitivity
  analysis}.{\BBCQ}
\newblock
\APACjournalVolNumPages{Wilmott}{2015}{78}{46--70}.
\newblock
\begin{APACrefDOI} \doi{10.1002/wilm.10434} \end{APACrefDOI}
\PrintBackRefs{\CurrentBib}

\bibitem [\protect \citeauthoryear {%
{BRODA Ltd}%
}{%
{BRODA Ltd}%
}{%
{\protect \APACyear {2022}}%
}]{%
BRODA}
\APACinsertmetastar {%
BRODA}%
\begin{APACrefauthors}%
{BRODA Ltd}.%
\end{APACrefauthors}%
\unskip\
\newblock
\APACrefYearMonthDay{2022}{}{}.
\newblock
{\BBOQ}\APACrefatitle {High-dimensional {S}obol' sequence generators}
  {High-dimensional {S}obol' sequence generators}.{\BBCQ}
\newblock
\APACjournalVolNumPages{http://www.broda.co.uk/}{}{}{}.
\PrintBackRefs{\CurrentBib}

\bibitem [\protect \citeauthoryear {%
Gatheral%
}{%
Gatheral%
}{%
{\protect \APACyear {2011}}%
}]{%
gatheral2011volatility}
\APACinsertmetastar {%
gatheral2011volatility}%
\begin{APACrefauthors}%
Gatheral, J.%
\end{APACrefauthors}%
\unskip\
\newblock
\APACrefYear{2011}.
\newblock
\APACrefbtitle {The volatility surface: a practitioner's guide} {The volatility
  surface: a practitioner's guide}.
\newblock
\APACaddressPublisher{}{John Wiley \& Sons}.
\newblock
\begin{APACrefDOI} \doi{10.1002/9781119202073} \end{APACrefDOI}
\PrintBackRefs{\CurrentBib}

\bibitem [\protect \citeauthoryear {%
Glasserman%
}{%
Glasserman%
}{%
{\protect \APACyear {2004}}%
}]{%
glasserman2004monte}
\APACinsertmetastar {%
glasserman2004monte}%
\begin{APACrefauthors}%
Glasserman, P.%
\end{APACrefauthors}%
\unskip\
\newblock
\APACrefYear{2004}.
\newblock
\APACrefbtitle {Monte Carlo methods in financial engineering} {Monte carlo
  methods in financial engineering}\ (\BVOL~53).
\newblock
\APACaddressPublisher{}{Springer}.
\newblock
\begin{APACrefDOI} \doi{10.1007/978-0-387-21617-1} \end{APACrefDOI}
\PrintBackRefs{\CurrentBib}

\bibitem [\protect \citeauthoryear {%
Jackel%
}{%
Jackel%
}{%
{\protect \APACyear {2008}}%
}]{%
Jac08}
\APACinsertmetastar {%
Jac08}%
\begin{APACrefauthors}%
Jackel, P.%
\end{APACrefauthors}%
\unskip\
\newblock
\APACrefYearMonthDay{2008}{}{}.
\newblock
{\BBOQ}\APACrefatitle {Hyperbolic local volatility} {Hyperbolic local
  volatility}.{\BBCQ}
\newblock
\APACjournalVolNumPages{Working paper available in
  http://www.jaeckel.org/HyperbolicLocalVolatility.pdf}{}{}{}.
\PrintBackRefs{\CurrentBib}

\bibitem [\protect \citeauthoryear {%
Kloeden%
\ \BBA {} Platen%
}{%
Kloeden%
\ \BBA {} Platen%
}{%
{\protect \APACyear {2013}}%
}]{%
Kloeden:Platen91}
\APACinsertmetastar {%
Kloeden:Platen91}%
\begin{APACrefauthors}%
Kloeden, P\BPBI E.%
\BCBT {}\ \BBA {} Platen, E.%
\end{APACrefauthors}%
\unskip\
\newblock
\APACrefYear{2013}.
\newblock
\APACrefbtitle {Numerical solution of stochastic differential equations}
  {Numerical solution of stochastic differential equations}.
\newblock
\APACaddressPublisher{}{Springer Science \& Business Media}.
\newblock
\begin{APACrefDOI} \doi{10.1007/978-3-662-12616-5} \end{APACrefDOI}
\PrintBackRefs{\CurrentBib}

\bibitem [\protect \citeauthoryear {%
L'Ecuyer%
}{%
L'Ecuyer%
}{%
{\protect \APACyear {2018}}%
}]{%
Ecuyer18}
\APACinsertmetastar {%
Ecuyer18}%
\begin{APACrefauthors}%
L'Ecuyer, P.%
\end{APACrefauthors}%
\unskip\
\newblock
\APACrefYearMonthDay{2018}{}{}.
\newblock
{\BBOQ}\APACrefatitle {Randomized quasi-Monte Carlo: An introduction for
  practitioners} {Randomized quasi-monte carlo: An introduction for
  practitioners}.{\BBCQ}
\newblock
\BIn{} J.~Fagerberg, D\BPBI C.~Mowery\BCBL {}\ \BBA {} R\BPBI R.~Nelson\
  (\BEDS), \APACrefbtitle {{Monte Carlo and Quasi-Monte Carlo Methods}} {{Monte
  Carlo and Quasi-Monte Carlo Methods}}\ (\BVOL~241, \BPGS\ 29--52).
\newblock
\APACaddressPublisher{}{Springer}.
\newblock
\begin{APACrefDOI} \doi{10.1007/978-3-319-91436-7_2} \end{APACrefDOI}
\PrintBackRefs{\CurrentBib}

\bibitem [\protect \citeauthoryear {%
Owen%
}{%
Owen%
}{%
{\protect \APACyear {1997}}%
}]{%
owen1997scrambled}
\APACinsertmetastar {%
owen1997scrambled}%
\begin{APACrefauthors}%
Owen, A\BPBI B.%
\end{APACrefauthors}%
\unskip\
\newblock
\APACrefYearMonthDay{1997}{}{}.
\newblock
{\BBOQ}\APACrefatitle {Scrambled net variance for integrals of smooth
  functions} {Scrambled net variance for integrals of smooth functions}.{\BBCQ}
\newblock
\APACjournalVolNumPages{The Annals of Statistics}{25}{4}{1541--1562}.
\newblock
\begin{APACrefDOI} \doi{10.1214/aos/1031594731} \end{APACrefDOI}
\PrintBackRefs{\CurrentBib}

\bibitem [\protect \citeauthoryear {%
Renzitti%
, Bastani%
\BCBL {}\ \BBA {} Sivorot%
}{%
Renzitti%
\ \protect \BOthers {.}}{%
{\protect \APACyear {2020}}%
}]{%
renzitti2020}
\APACinsertmetastar {%
renzitti2020}%
\begin{APACrefauthors}%
Renzitti, S.%
, Bastani, P.%
\BCBL {}\ \BBA {} Sivorot, S.%
\end{APACrefauthors}%
\unskip\
\newblock
\APACrefYearMonthDay{2020}{}{}.
\newblock
{\BBOQ}\APACrefatitle {Accelerating CVA and CVA Sensitivities Using Quasi-Monte
  Carlo Methods} {Accelerating cva and cva sensitivities using quasi-monte
  carlo methods}.{\BBCQ}
\newblock
\APACjournalVolNumPages{Wilmott}{2020}{108}{78--93}.
\newblock
\begin{APACrefDOI} \doi{10.2139/ssrn.3193219} \end{APACrefDOI}
\PrintBackRefs{\CurrentBib}

\bibitem [\protect \citeauthoryear {%
Sobol'%
, Asotsky%
, Kreinin%
\BCBL {}\ \BBA {} Kucherenko%
}{%
Sobol'%
\ \protect \BOthers {.}}{%
{\protect \APACyear {2011}}%
}]{%
SobAsoKreiKuch11}
\APACinsertmetastar {%
SobAsoKreiKuch11}%
\begin{APACrefauthors}%
Sobol', I\BPBI M.%
, Asotsky, D.%
, Kreinin, A.%
\BCBL {}\ \BBA {} Kucherenko, S.%
\end{APACrefauthors}%
\unskip\
\newblock
\APACrefYearMonthDay{2011}{}{}.
\newblock
{\BBOQ}\APACrefatitle {Construction and comparison of high-dimensional {S}obol'
  generators} {Construction and comparison of high-dimensional {S}obol'
  generators}.{\BBCQ}
\newblock
\APACjournalVolNumPages{Wilmott}{2011}{56}{64--79}.
\newblock
\begin{APACrefDOI} \doi{10.1002/wilm.10056} \end{APACrefDOI}
\PrintBackRefs{\CurrentBib}

\end{thebibliography}

\end{document}